\documentclass[twocolumn,floatfix,aps,superscriptaddress,eqsecnum]{revtex4}
\usepackage{graphicx}
\usepackage{amsmath}
\usepackage{amssymb}
\usepackage{bm}
      
\begin{document}

\title{Weyl-Majorana solenoid}
\author{P. Baireuther}
\affiliation{Instituut-Lorentz, Universiteit Leiden, P.O. Box 9506, 2300 RA Leiden, The Netherlands}
\author{J. Tworzyd{\l}o}
\affiliation{Institute of Theoretical Physics, Faculty of Physics, University of Warsaw, ul.\ Pasteura 5, 02--093 Warszawa, Poland}
\author{M. Breitkreiz}
\affiliation{Instituut-Lorentz, Universiteit Leiden, P.O. Box 9506, 2300 RA Leiden, The Netherlands}
\author{\.{I}. Adagideli}
\affiliation{Faculty of Engineering and Natural Sciences, Sabanci University, Orhanli-Tuzla, 34956, Istanbul, Turkey}
\author{C. W. J. Beenakker}
\affiliation{Instituut-Lorentz, Universiteit Leiden, P.O. Box 9506, 2300 RA Leiden, The Netherlands}
\date{December 2016}
\begin{abstract}
A Weyl semimetal wire with an axial magnetization has metallic surface states (Fermi arcs) winding along its perimeter, connecting bulk Weyl cones of opposite topological charge (Berry curvature). We investigate what happens to this ``Weyl solenoid'' if the wire is covered with a superconductor, by determining the dispersion relation of the surface modes propagating along the wire. Coupling to the superconductor breaks up the Fermi arc into a pair of Majorana modes, separated by an energy gap. Upon variation of the coupling strength along the wire there is a gap inversion that traps the Majorana fermions.
\end{abstract} 
\maketitle

\section{Introduction}

A three-dimensional Weyl semimetal has topological features that are lacking in its two-dimensional counterpart, graphene \cite{Nie83,Vol03,Rao16}. One striking feature is the appearance of surface states, in Fermi arcs connecting Weyl cones of opposite topological charge (Chern number or Berry curvature) \cite{Wan11}. Unlike the surface states of a topological insulator, which are the only source of metallic conduction, the Fermi arcs at the surface compete with the Weyl cones in the bulk when it comes to transport properties. Quantum oscillations in the magnetoresistance are one example of an effect where the Fermi arcs play a prominent role \cite{Pot14,Zha16}, the chiral magnetic effect without Landau levels is another example \cite{Bai16}.

\begin{figure}[tb]
\centerline{\includegraphics[width=0.8\linewidth]{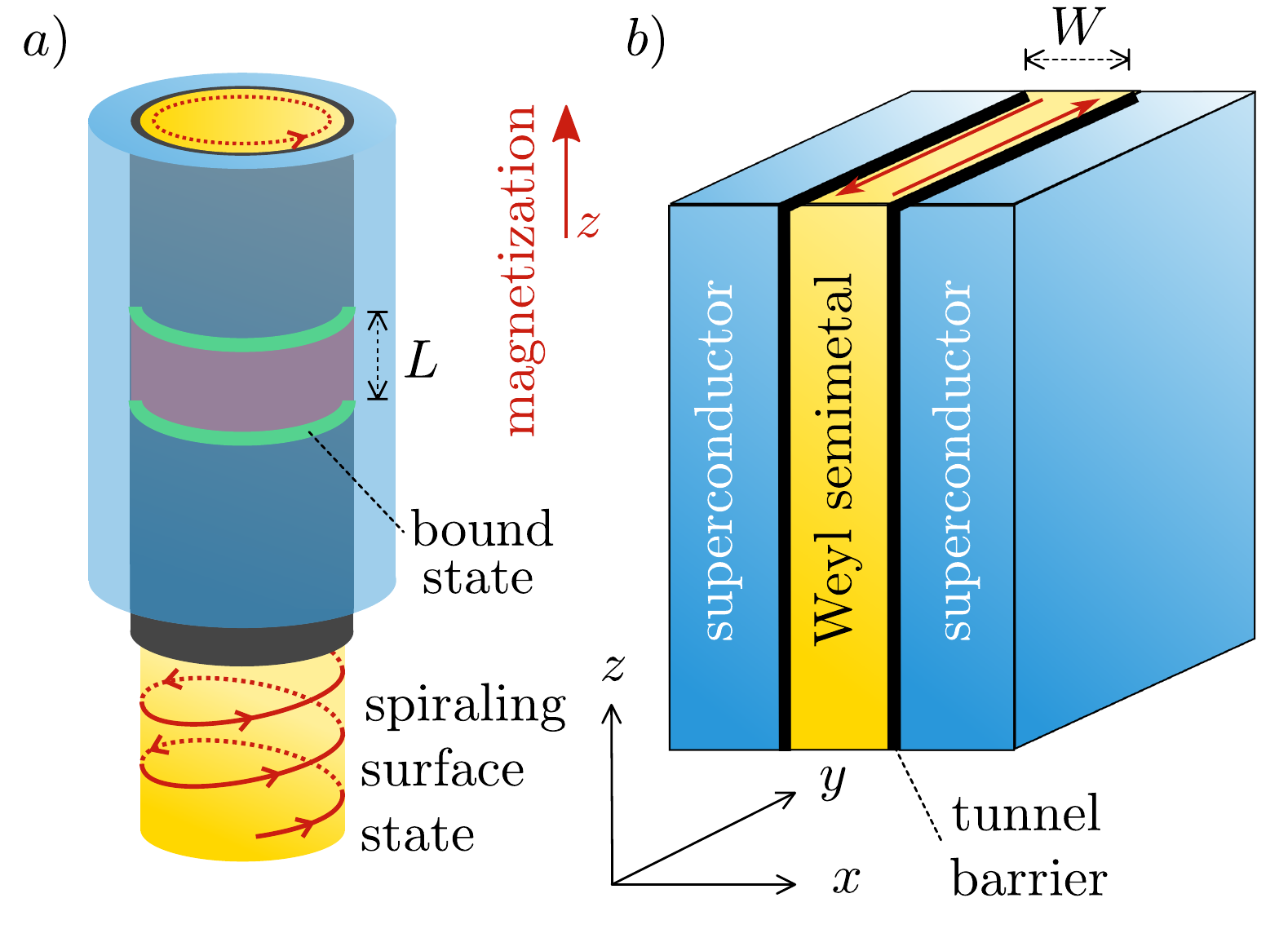}}
\caption{Panel a) Weyl-Majorana solenoid, formed by a Weyl semimetal wire with an axial magnetization, coupled via a tunnel barrier to a superconductor. Charge-neutral Majorana modes propagate along the wire, confined to the normal-superconductor (NS) interface. A gap inversion in a segment of length $L$, induced by a variation in coupling strength, traps a pair of quasiparticles at the two ends of the segment. Panel b) SNS slab geometry to study the Majorana modes at the NS interface.
}
\label{fig_layout}
\end{figure}

An interesting way to differentiate surface from bulk is to bring the Weyl semimetal into contact with a superconductor. While the Weyl cones in the bulk remain largely unaffected, the surface states acquire the mixed electron-hole character of a charge-neutral Bogoliubov quasiparticle --- a Majorana fermion \cite{Men12,Che13,Uch14,Kha14,Kha16,Che16}. Here we investigate this proximity effect in the nanowire geometry of Fig.\ \ref{fig_layout}, in which an axial magnetization causes the surface modes to spiral along the wire, essentially forming a solenoid on the nanoscale \cite{Bai16}. We study the dispersion relation of the Majorana modes and identify a mechanism to trap the quasiparticles at a specified location along the wire.

In the next section we identify the pair of $\mathbb{Z}_2$ quantum numbers $\nu,\kappa$ that label the four surface modes in a given orbital subband. The electron-hole index $\nu$ is generic for any surface state where electrons and holes are coupled by Andreev reflection \cite{Hop00,Rak07,Ost11}. The connectivity index $\kappa$ is specific for the Fermi arcs, it distinguishes whether the surface state reconnects in the bulk with the Weyl cone at positive or negative energy. In Sec.\ \ref{sec_effectiveH} we construct the $4\times 4$ matrix Hamiltonian in the $\nu,\kappa$ basis, constrained by particle-hole symmetry, as an effective low-energy description of the two-dimensional surface modes. 

We then proceed in Sec.\ \ref{sec_micro} with a numerical calculation of the three-dimensional band structure of a microscopic model Hamiltonian. The unexpected feature revealed by this simulation is a gap inversion, visible in the band structure as a level crossing between two surface modes with the same connectivity index. The gap inversion can be controlled by variation of the tunnel coupling between the semimetal and the superconductor. At the domain wall where the gap changes sign, a charge-neutral quasiparticle is trapped --- as we demonstrate numerically and explain within the context of the effective surface Hamiltonian in Sec.\ \ref{sec_trapping}. In Sec.\ \ref{analytics} we study the same gap inversion analytically, via a mode-matching calculation. In the concluding Sec.\ \ref{sec_conclude} we comment on the relation of the gap inversion to the flow of Berry curvature in the Brillouin zone.

\section{Connectivity index of surface Fermi arcs}
\label{sec_connect}

The geometry under consideration is shown in Fig.\ \ref{fig_layout}. A Weyl semimetal wire oriented along the $z$-axis is covered by a superconductor. We include a thin insulating layer between the superconductor and the Weyl semimetal, forming a tunnel barrier. A magnetization in the $z$-direction breaks time-reversal symmetry and separates the Weyl cones along the $p_z$ momentum direction in the Brillouin zone. (Induced superconductivity in the presence of time-reversal symmetry, with minimally four Weyl points, has a different phenomenology \cite{Che16}.) The surface states connecting the Weyl cones are chiral, circulating with velocity $v_\phi$ in a direction set by the magnetization. If inversion symmetry is broken the surface states also spiral with velocity $v_z$ along the wire \cite{Bai16}.

\begin{figure}[tb]
\centerline{\includegraphics[width=1\linewidth]{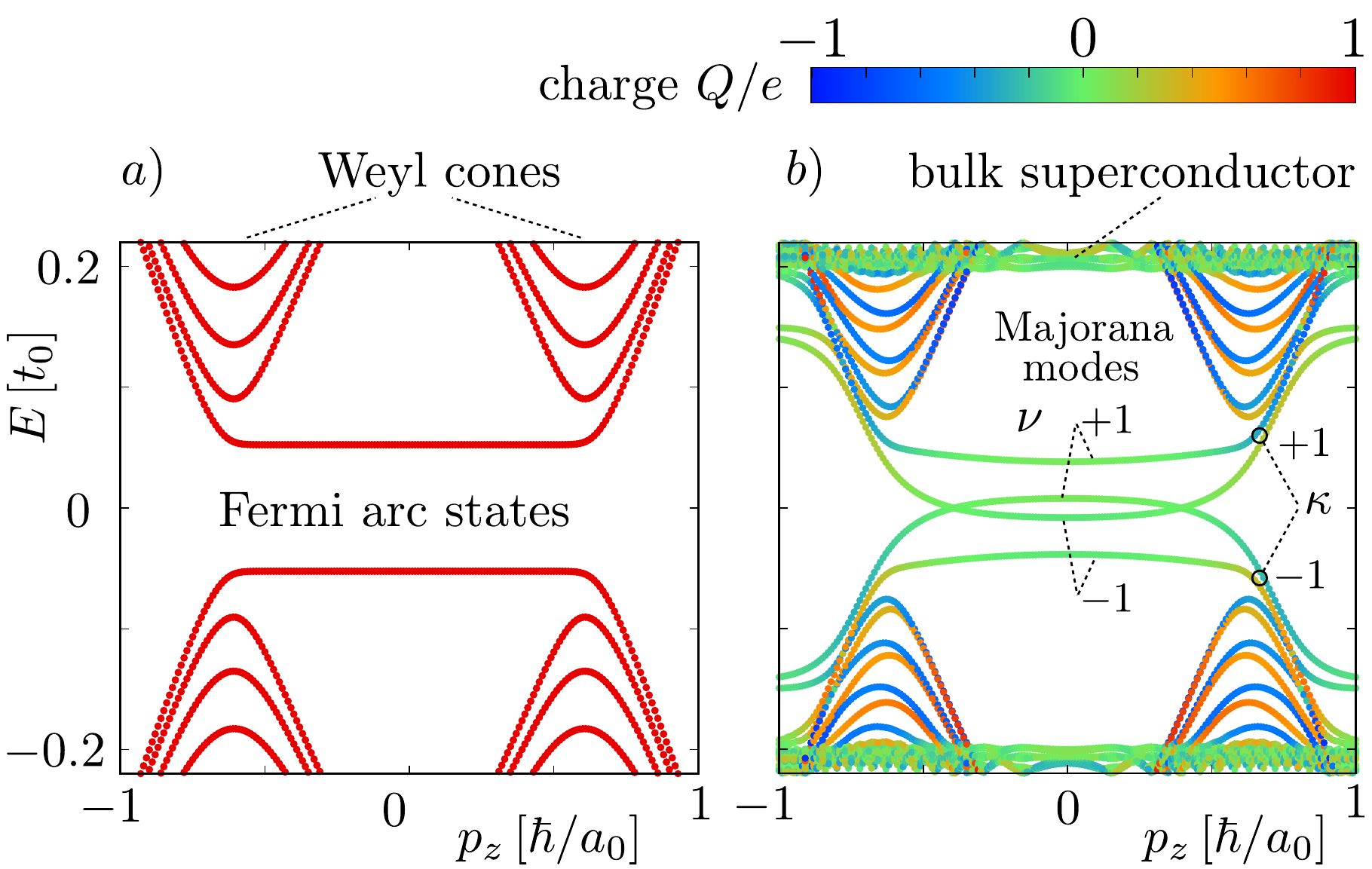}}
\caption{Band structure of a Weyl semimetal in the slab geometry of Fig.\ \ref{fig_layout}b, calculated from the tight-binding model described in the text \cite{parameters_num1}. In panel a) there is only the Weyl semimetal, in panel b) the superconducting contacts have been added. Inversion symmetry has not been broken, so the spectrum has $\pm p_z$ symmetry, in addition to the particle-hole symmetry $E(p_z)=-E(-p_z)$. In the slab geometry the transverse wave vector $k_y$ is a good quantum number, and to make the figure less crowded only subbands at a single value of $k_y$ are shown. (The Fermi arcs in panel a) are approximately at $\pm v_{\phi}\sin k_y$.) The superconductor breaks up the two Dirac fermion surface modes in panel a) into four Majorana fermion modes in panel b), labeled by a pair of indices $\kappa,\nu=\pm 1$. The Majorana modes are nearly charge-neutral, as indicated by the color scale (with electron charge $+e$).
}
\label{fig_numerics1}
\end{figure}

As shown in Fig.\ \ref{fig_numerics1}, resulting from a model calculation described in Sec.\ \ref{sec_micro}, at the interface with a  superconductor the surface spectrum is drastically modified. We seek an effective Hamiltonian that describes this proximity effect on the Fermi arcs.

The first question we have to address is which pairs of states are coupled by the superconducting pair potential $\Delta$. In the bulk spectrum the answer is well known \cite{Men12,Kha16}: Superconductivity couples electrons in a Weyl cone of positive Berry curvature to holes in a Weyl cone of negative Berry curvature, and vice versa. To decide this question for the surface states, we assign to each Fermi arc a ``connectivity index'' $\kappa=\pm 1$, depending on whether it reconnects in the bulk with the Weyl cone at positive or negative energy. Inspection of Fig.\ \ref{fig_numerics1} shows that $\Delta$ predominantly couples Fermi arcs with same $\kappa$, pushing them apart, without removing the crossing between states of opposite $\kappa$.

More explicitly, in a slab geometry we can identify $\kappa={\rm sign}\,k_y$ and in a cylindrical wire geometry we would have $\kappa={\rm sign}\,p_\phi$. The coupling of states with different $\kappa$ is then forbidden by (translational or rotational) symmetry. More generally, in the absence of any symmetry, the sign of $\kappa=\pm 1$ says whether the Fermi arc connects with the Weyl cone at $\pm E$, and thus identifies which pairs of Fermi arcs are predominantly coupled by $\Delta$.

\section{Effective surface Hamiltonian}
\label{sec_effectiveH}

The superconducting proximity effect is governed by the Bogoliubov-De Gennes (BdG) Hamiltonian, describing the coupling of electrons and holes by the pair potential. In the numerical simulations we will work with the BdG Hamiltonian in a 3D microscopic model. For analytical insight we aim for an effective 2D description involving only surface modes.

Each orbital subband $n$ is associated with four Majorana modes, labeled by a pair of $\mathbb{Z}_2$ indices $\kappa,\nu$. (See Fig.\ \ref{fig_numerics1}.) The connectivity index $\kappa=\pm$ identifies the connectivity of the surface mode (with the Weyl cone at positive or negative energy), the electron-hole index $\nu=\pm$ identifies the pair of Majorana fermions that form a Dirac fermion.  The corresponding BdG Hamiltonian ${\cal H}_n$ is a $4\times 4$ matrix with $p_z$-dependent elements. In what follows we omit the subband index $n$ for ease of notation. 

The fundamental symmetry of the BdG Hamiltonian is particle-hole symmetry,
\begin{equation}
{\cal H}(p_z)=-\kappa_y\nu_y{\cal H}^\ast(-p_z)\kappa_y\nu_y,\label{Cphsymdef}
\end{equation}
with Pauli matrices $\kappa_\alpha$ and $\nu_\alpha$ acting, respectively on the connectivity and electron-hole degree of freedom ($\alpha=1,2,3\mapsto x,y,z$ and $\alpha=0$ for the unit matrix). The operation of particle-hole conjugation squares to $+1$, which places the system in symmetry class D \cite{Ryu10} --- this is the appropriate symmetry class in the absence of time-reversal and spin-rotation symmetry.

If we neglect the mixing by disorder of surface states with opposite connectivity index $\kappa=\pm$, the $4\times 4$ matrix ${\cal H}$ decouples into two blocks $H_\pm$ related by particle-hole symmetry,
\begin{equation}
{\cal H}=\begin{pmatrix}
H_+&0\\
0&H_-
\end{pmatrix},\;\;
H_-(p_z)=-\nu_y H_+^\ast(-p_z)\nu_y.
\label{HHplusHminus}
\end{equation}
The $2\times 2$ matrices $H_\pm$ can be decomposed into Pauli matrices,
\begin{equation}
H_\pm(p_z)=\pm D_0(\pm p_z)\nu_0+\textstyle{\sum_{\alpha=1}^3}D_\alpha(\pm p_z)\nu_\alpha,\label{HDdecomposition}
\end{equation}
with real $p_z$-dependent coefficients $D_\alpha$.

Diagonalization of the Hamiltonian \eqref{HHplusHminus} gives the dispersion relation $E_{\kappa,\nu}(p_z)$ of the four Majorana modes in the $n$-th subband,
\begin{equation}
E_{\kappa,\nu}(p_z)=\kappa D_0(\kappa p_z) +\nu\sqrt{\textstyle{\sum_{\alpha=1}^3} D_\alpha^2(\kappa p_z)}.
\label{Enpzresult}
\end{equation}
Particle-hole symmetry is expressed by $E_{\kappa,\nu}(p_z)=-E_{-\kappa,-\nu}(-p_z)$. Inversion symmetry, $E_{\kappa,\nu}(p_z)=E_{\kappa,\nu}(-p_z)$, is satisfied if $D_0$ is an even function of $p_z$ while each of the functions $D_1,D_2,D_3$ has a definite parity (even or odd).

\section{Numerical simulation of a microscopic model}
\label{sec_micro}

We now turn to a microscopic model of a Weyl semimetal in contact with a superconductor, which we solve numerically. The Weyl semimetal has BdG Hamiltonian
\begin{align}
&H_{\rm W}(\bm{k})=\nu_z\tau_z(t\sigma_x\sin k_x+t\sigma_y\sin k_y +t_z \sigma_z\sin k_z)\nonumber\\
&\quad+{m}(\bm{k})\nu_z\tau_x\sigma_0+\lambda\nu_z\tau_z\sigma_0+\beta \nu_0\tau_0 \sigma_z
-\mu \nu_z\tau_0\sigma_0,\nonumber\\
&{m}(\bm{k})={m}_0+t(2-\cos k_x-\cos k_y)+t_z(1-\cos k_z),\label{Hsymmetric}
\end{align}
with chemical potential $\mu$ and charge operator
\begin{equation}
Q=-e\frac{\partial H_{\rm W}}{\partial \mu}=e\nu_z\tau_0\sigma_0.\label{QWdef}
\end{equation}
\begin{figure}[tb]
\centerline{\includegraphics[width=.8\linewidth]{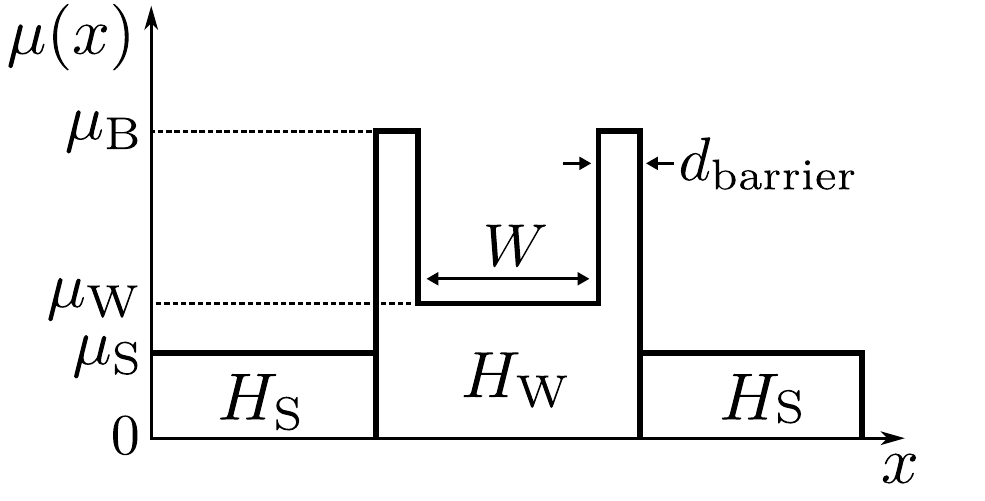}}
\caption{Spatial profile of the chemical potential $\mu(x)$.}
\label{fig_layout_2}
\end{figure}
The Pauli matrices $\sigma_\alpha$ and $\tau_\alpha$ refer to spin and orbital degrees of freedom, respectively, while $\nu_\alpha$ acts on the electron-hole index. The momentum $\bm{k}$ varies over the Brillouin zone $|k_\alpha| < \pi$ of a simple cubic lattice (lattice constant $a_0\equiv 1$). This is a model of a layered material in the Bi$_2$Se$_3$ family \cite{Vaz13}, with weak coupling $t_z<t$ in the $z$-direction, perpendicular to the layers in the $x$--$y$ plane.

The particle-hole symmetry relation is
\begin{equation}
H_{\rm W}(\bm{k})=-\sigma_y\nu_yH_{\rm W}^\ast(-\bm{k})\sigma_y\nu_y.\label{Hkphsym}
\end{equation}
The magnetization term $\propto\beta$ breaks time-reversal symmetry, $H_{\rm W}(\bm{k})=\sigma_yH_{\rm W}^\ast(-\bm{k})\sigma_y$. Inversion symmetry, $H_{\rm W}(\bm{k})=\tau_xH_{\rm W}(-\bm{k})\tau_x$, is broken by the strain term $\propto \lambda$.

The Weyl semimetal is in contact with a spin-singlet \textit{s}-wave superconductor, with Hamiltonian
\begin{align}
H_{\rm S}={}&[\tilde{t}(2-\cos k_x-\cos k_y)+\tilde{t}_z(1-\cos k_z)]\nu_z\tau_0\sigma_0\nonumber\\
&-\mu \nu_z\tau_0\sigma_0+\Delta_0\nu_x\tau_0\sigma_0.\label{HSdef}
\end{align}
There are different chemical potentials in the Weyl semimetal, $\mu=\mu_{\rm W}$, and in the superconductor, $\mu=\mu_{\rm S}$. At the NS interface we include an electrostatic potential barrier of width $d_{\rm{barrier}}$, raising $\mu$ to a value $\mu_{\rm B}\equiv U_{\rm barrier}$. The resulting spatial profile $\mu(x)$ is shown in Fig.\ \ref{fig_layout_2}.
 
We consider the two geometries shown in Fig.\ \ref{fig_layout}, a wire geometry and a computationally more efficient slab geometry \cite{kwant}. In each case there is translational invariance along the $z$-direction. In the slab geometry there is in addition translational invariance in the $y$-direction, so the modes are labeled by a continuous quantum number $k_y$ \cite{note1}.

\begin{figure}[tb]
\centerline{\includegraphics[width=1\linewidth]{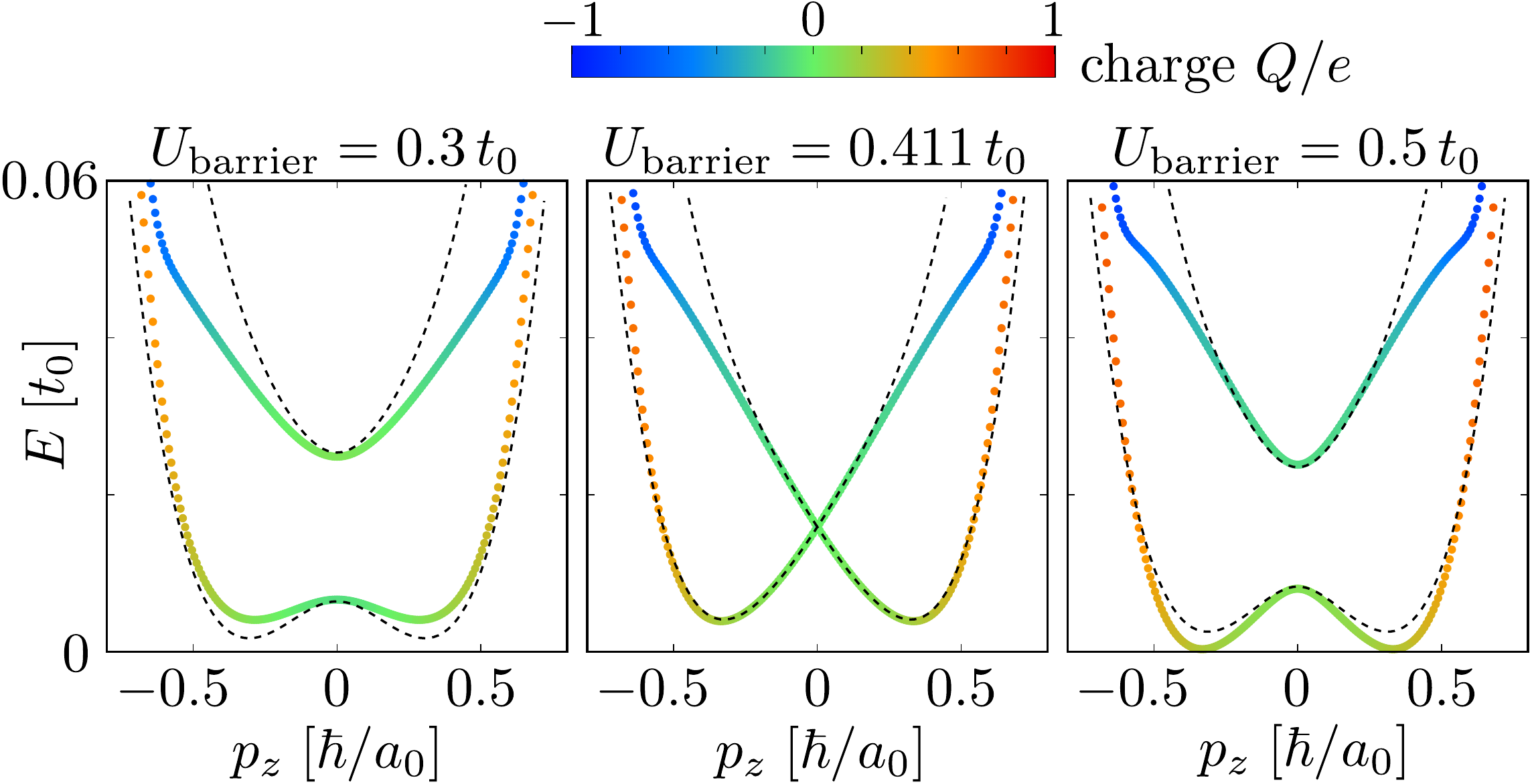}}
\caption{Data points: Band structure in the slab geometry (colored according to the charge expectation value), showing the level crossing at $p_z=0$ between a pair of Majorana modes with $\kappa=+1$, $\nu=\pm 1$. The parameters are those of Fig.\ \ref{fig_numerics1}b \cite{parameters_num1}, except for the tunnel barrier height $U_{\rm barrier}$, which is varied to tune through the gap inversion. The dashed curves are fits \cite{parameters_fit} to the dispersion \eqref{Ecrossingmodel} from the effective surface Hamiltonian.
}
\label{fig_numerics_2}
\end{figure}

\begin{figure}[tb]
\centerline{\includegraphics[width=1\linewidth]{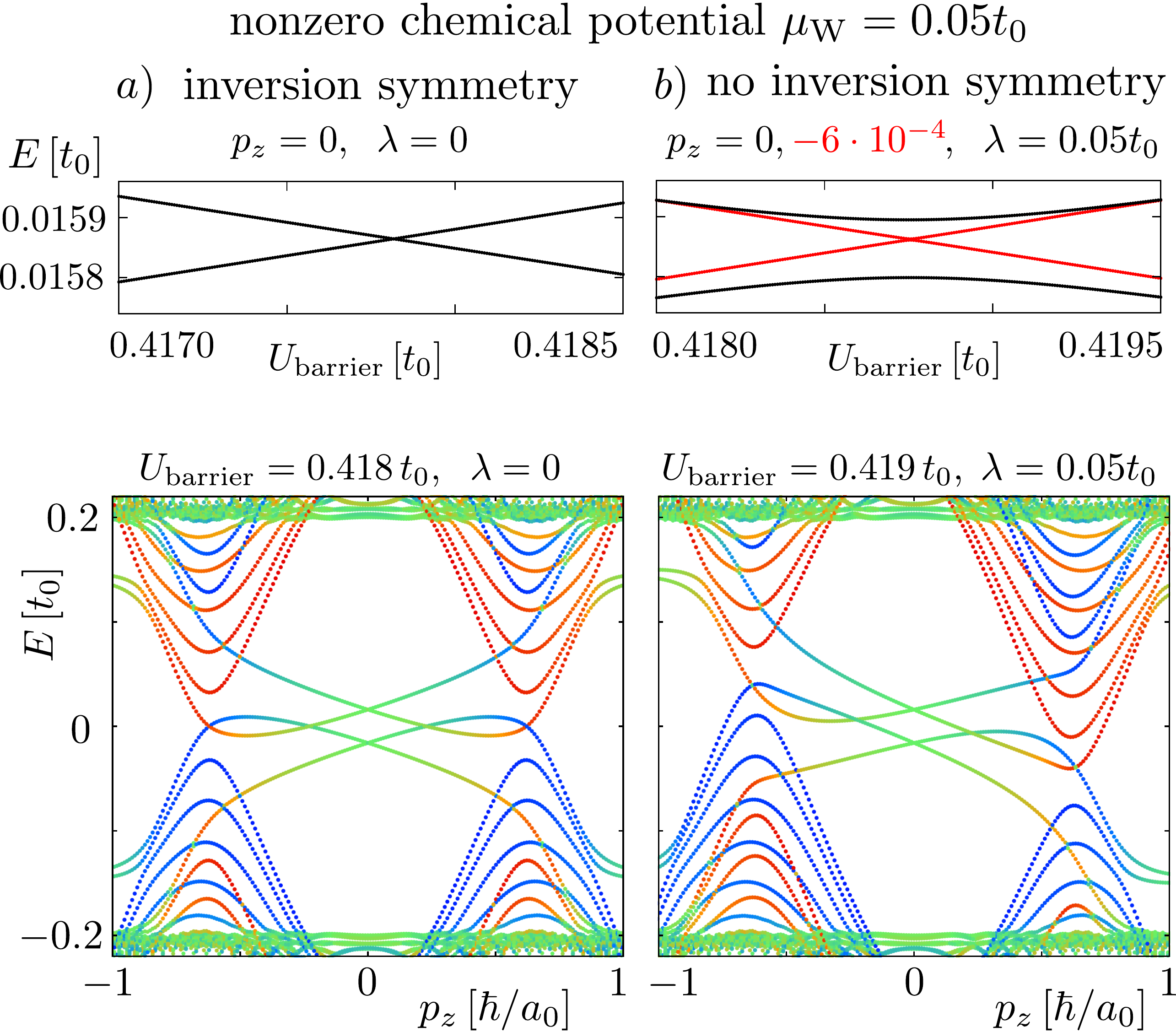}}
\caption{Band structure in the slab geometry, showing the level crossing near $p_z=0$ between modes with the same connectivity index. In the lower panels we show the crossing as a function of $p_z$ at fixed tunnel barrier height $U_{\rm barrier}$, in the upper panels we show the crossing at fixed $p_z$ as a function of $U_{\rm barrier}$. The parameters and color scale are those of Fig.\ \ref{fig_numerics1}b \cite{parameters_num1}, but we took a nonzero $\mu_{\rm W}=0.05\,t_0$ (notice the displacement of electron and hole bands in the bulk Weyl cones) in order to demonstrate that the level crossing does not require a vanishing chemical potential. The level crossing also persists if inversion symmetry is broken by a nonzero $\lambda=0.05\,t_0$, but the crossing point is displaced away from $p_z=0$ (compare black and red curves in panel b, at $p_z=0$ and $p_z=-6\cdot 10^{-4}\,\hbar/a_0$).
}
\label{fig_numerics_3}
\end{figure}

\begin{figure}[tb]
\centerline{\includegraphics[width=1\linewidth]{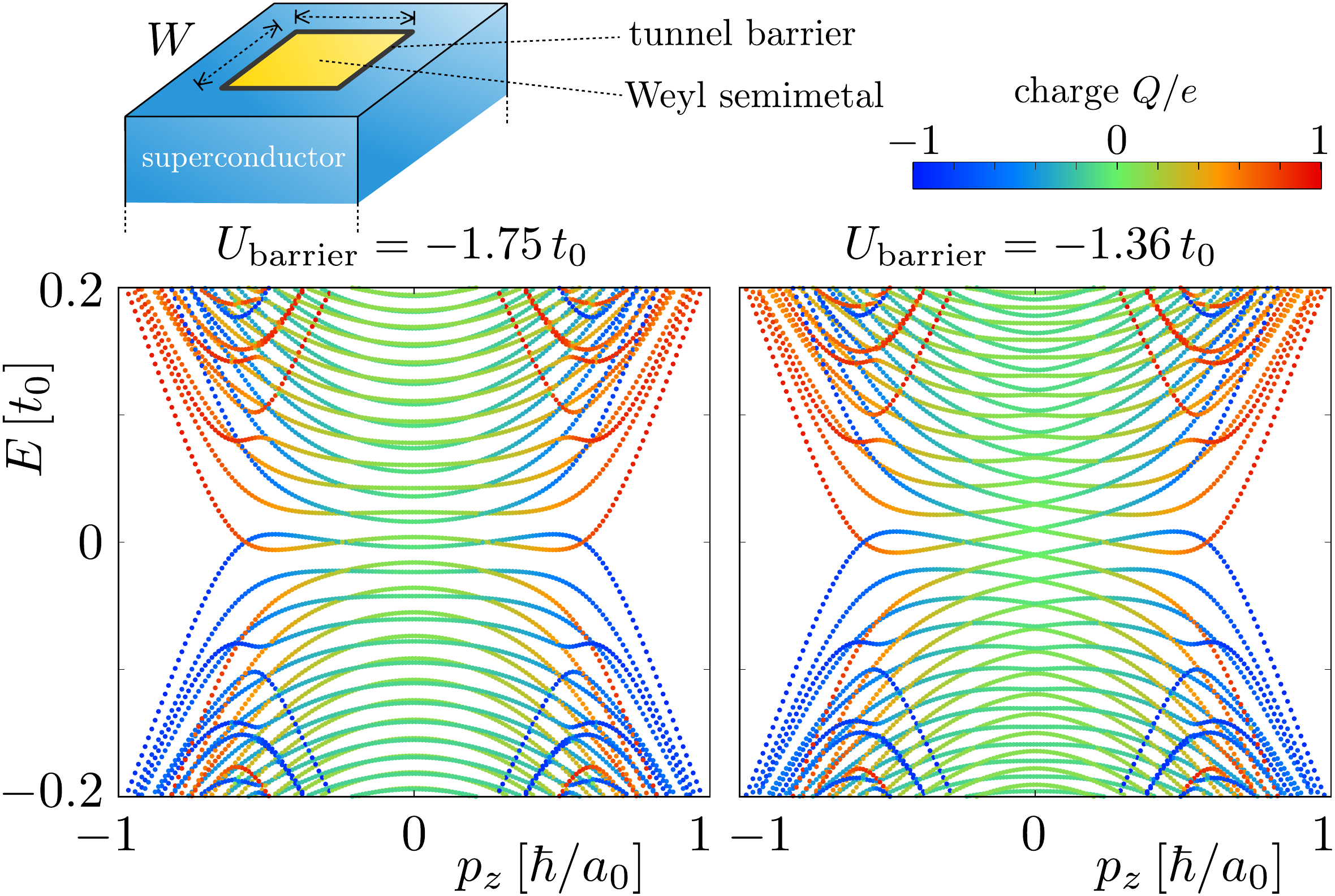}}
\caption{Band structure in a wire geometry (square cross section \cite{parameters_num4}), showing all modes in the energy range $-0.2<E/t_0<0.2$. (The previous plots in the slab geometry showed only the modes with a single $k_y$ value, but in the wire $k_y$ is not a good quantum number.) The gap between pairs of modes in the same subband and with the same connectivity index closes at $p_z=0$ upon variation of the tunnel barrier height.
}
\label{fig_numerics_4}
\end{figure}

The dispersion relation in the slab geometry is shown in Fig.\ \ref{fig_numerics1}. The mode crossings at nonzero $p_z$ appear because modes with different connectivity index $\kappa$ are uncoupled in the absence of disorder. In Fig.\ \ref{fig_numerics_2} we show a different type of crossing, near $p_z=0$ between modes with the same $\kappa$, induced by variation of the tunnel barrier height. This crossing appears generically when we vary interface parameters, in Fig.\ \ref{fig_numerics_3} we show that it persists at nonzero chemical potential $\mu=\mu_{\rm W}$ in the Weyl semimetal \cite{note_EF}. Inversion symmetry breaking by a nonzero $\lambda$ moves the crossing point away from $p_z=0$, but does not destroy it. The wire geometry gives similar results, see Fig.\ \ref{fig_numerics_4}.

To model this effect in the framework of the surface Hamiltonian \eqref{HDdecomposition}, we take a momentum-independent complex off-diagonal potential $D_1-iD_2\equiv\Delta$ with amplitude $\Delta_0=c(U_{\rm barrier}-U_c)$ that crosses zero at some critical barrier height $U_c$. Inversion symmetry imposes a definite parity on the real diagonal potential $D_3\equiv\mu(p_z)$, such that even a small admixture of an odd-parity component enforces $\mu(0)=0$ when $\lambda=0$. If we take $\mu(p_z)=c' \lambda+ c'' p_z$ the dispersion relation \eqref{Enpzresult} in the pair of modes with $\kappa=+1$ has the form
\begin{equation}
E_\nu(p_z)=D_0(p_z) +\nu\sqrt{c^2(U_{\rm barrier}-U_c)^2+(c'\lambda+c''p_z)^2}.\label{Ecrossingmodel}
\end{equation}
The dashed curves in Fig.\ \ref{fig_numerics_2} are fits to this functional form, with $\lambda=0$ and a quartic $D_0(p_z)$. The qualitative behavior agrees reasonably well.

\section{Quasiparticle trapping by gap inversion}
\label{sec_trapping}

\begin{figure}[tb]
\centerline{\includegraphics[width=0.9\linewidth]{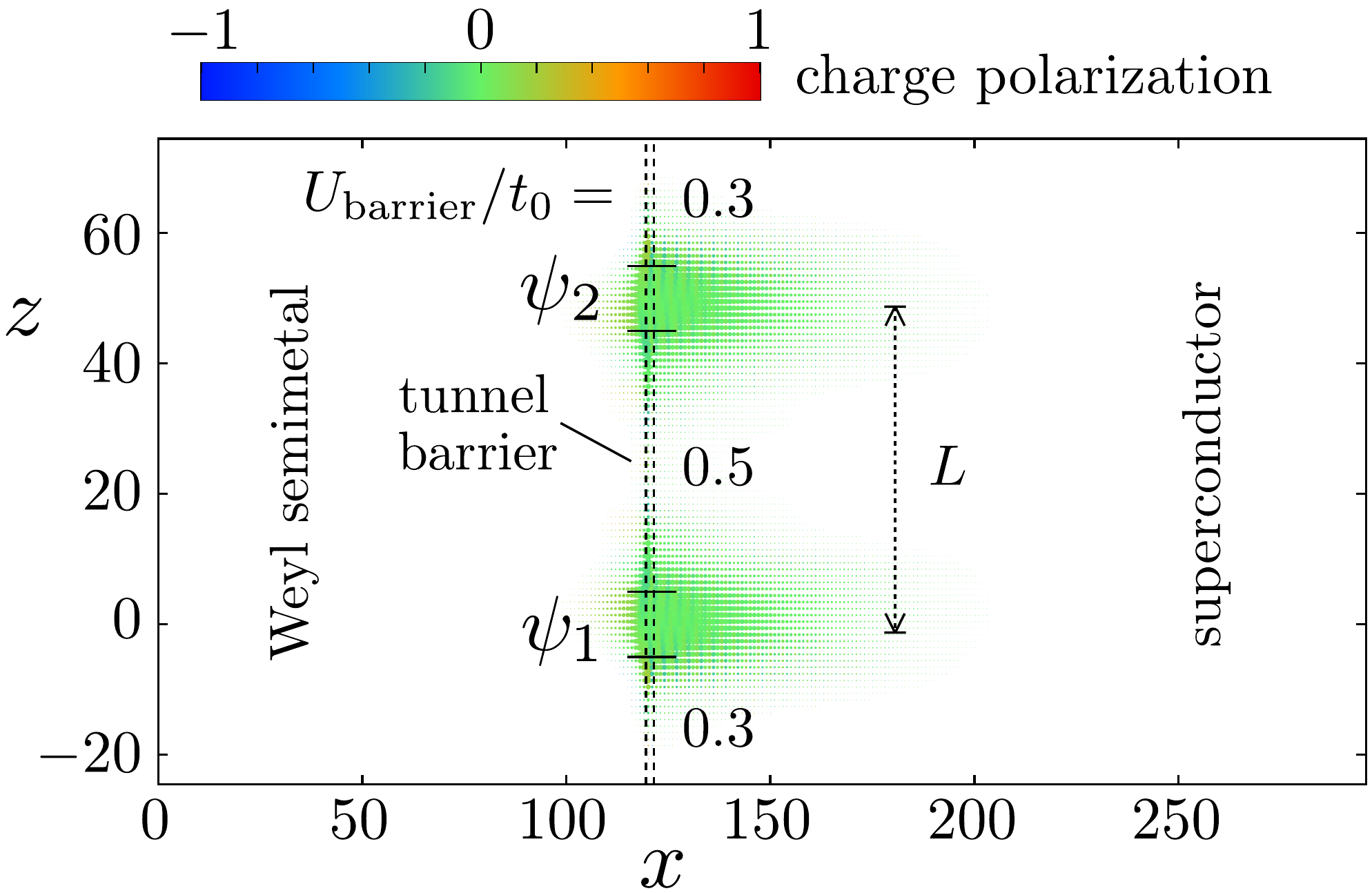}}   
\caption{Density of states (dot size) and local charge polarization (color) at $E= 0.016\,t_0$ in an NS junction in the slab geometry \cite{noteNSvsSNS} with a $z$-dependent tunnel barrier height. The vertical dashed lines indicate the tunnel barrier at the NS interface. The horizontal lines indicate the regions where the tunnel barrier height $U_{\rm barrier}$ is varied from $0.3\,t_0$ to $0.5\,t_0$ and back, passing through the critical value $U_c=0.411\,t_0$ near $z=0$ and $z=50\equiv L$. At these domain walls the gap between a pair of surface modes (at given $|k_y|=\pi/120$) closes and reopens, trapping a charge-neutral quasiparticle. The parameters are the same as in Fig.\ \ref{fig_numerics_2}, with periodic boundary conditions in the $z$-direction. } 
\label{fig_trapping}
\end{figure}

The gap inversion of Fig.\ \ref{fig_numerics_2} can be used to trap a quasiparticle by varying the tunnel barrier height $U_{\rm barrier}(z)$ (by means of a variation in the thickness of the insulating layer), from a value above the critical strength $U_c$ to a value below $U_c$. A demonstration of this effect in the slab geometry is shown in Fig.\ \ref{fig_trapping}, where we plot the local density of states and charge polarization $\langle \psi|\nu_z|\psi\rangle\langle\psi|\psi\rangle^{-1}\in(-1,+1)$ at each site of the lattice.

In terms of the surface Hamiltonian, the quasiparticle trapping is described by the Schr\"{o}dinger equation $H_\pm\psi(z)=E\psi(z)$ with
\begin{equation}
H_\pm=\begin{pmatrix}
\pm D_0(\pm p_z)+\mu(\pm p_z)&\Delta(z)\\
\Delta^\ast(z)&\pm D_0(\pm p_z)-\mu(\pm p_z)
\end{pmatrix}.\label{HpmmuDelta}
\end{equation}
We take a real $\Delta(z)=c(U_{\rm barrier}(z)-U_c)$ and, respectively, an even and odd $p_z$-dependence of $D_0$ and $\mu=c'' p_z$ --- consistent with inversion symmetry. If we neglect quadratic terms in $D_0$ we have a matrix differential equation of first order,
\begin{equation}
\mp i\hbar c''\nu_z\frac{d\psi}{dz}=\bigl[\bigl(E\mp D_0(0)\bigr)\nu_0-\Delta(z)\nu_x\bigr]\psi(z).\label{HpsiEpsi}
\end{equation}

Let $\Delta(z)/c''$ vary from a positive value for $z<0$ and $z>L$ to a negative value in the interval $0<z<L$. For sufficiently large $L$ we can consider the domain wall at $z=0$ separately from the one at $z=L$. At energy $E=\pm D_0(0)$ there is a bound state at $z=0$ with wave function
\begin{equation}
\psi_\pm(z)=\exp\left(\pm\frac{1}{\hbar c''}\int_0^z dz'\,\Delta(z')\nu_y\right)\psi_\pm(0).\label{uzprofile}
\end{equation}
This should be a decaying function of $|z|$, so $\psi_\pm(0)=(1,\pm i )$ is an eigenstate of $\nu_y$ with eigenvalue $\pm 1$.

Fig.\ \ref{fig_trapping} shows that the bound state is a charge-neutral quasiparticle. There is one state at energy $+D_0(0)$ and a second state at $-D_0(0)$, but because the BdG equation doubles the spectrum only a single Majorana fermion is trapped at $z=0$. A second Majorana fermion is trapped at $z=L$. All of this is for a single orbital mode $n$. We have found numerically that the critical barrier height $U_c$ is weakly $n$-dependent, so a domain wall traps one Majorana fermion per orbital subband.

\section{Analytical mode-matching calculation}
\label{analytics}

\subsection{Hamiltonian with spatially dependent coefficients}

To analytically substantiate our numerical findings we have performed a mode-matching calculation in the slab geometry of Fig.\ \ref{fig_layout}b, matching electron and hole modes in the normal (N) region $0<x<W$ to Bogoliubov quasiparticles in the superconducting (S) regions $x<0$, $x>W$. This procedure can be greatly simplified if we choose a single BdG Hamiltonian $H$ with $x$-dependent coefficients, rather than the different $H_{\rm W}$ and $H_{\rm S}$ of Sec.\ \ref{sec_micro} --- the former choice is a less realistic model of an SNS junction than the latter, but as we will see the results are essentially equivalent. 

Our starting point is therefore the Hamiltonian
\begin{align}
H={}&\nu_z\tau_z(t\sigma_x\sin k_x+t\sigma_y\sin k_y +t_z \sigma_z\sin k_z)\nonumber\\
&+m\nu_z\tau_x\sigma_0+\lambda\nu_z\tau_z\sigma_0+\beta \nu_0\tau_0 \sigma_z\nonumber\\
&-\mu(x)\nu_z\tau_0\sigma_0+\Delta(x)\nu_x\tau_0\sigma_0,\label{Hx} 
\end{align}
with chemical potential $\mu(x)$, pair potential $\Delta(x)$, and mass term
\begin{equation}
m(\bm{k})={m}_0+t(2-\cos k_x-\cos k_y)+t_z(1-\cos k_z).\label{mdef}
\end{equation}

We will compare our analytical mode-matching calculation to a numerical solution of the discretized Hamiltonian \eqref{Hx}. For this analytics, but not for the numerics, we make one further simplification, which is to linearize the Hamiltonian in the transverse momentum component $k_x$, so that the mode-matching calculation requires the solution of a set of first order differential equation in $x$. We thus replace $\sin k_x\mapsto k_x$ and replace the mass term \eqref{mdef} by
\begin{equation}
\tilde{m}(k_y,k_z)={m}_0+t(1-\cos k_y)+t_z(1-\cos k_z).\label{mdef2}
\end{equation}

\subsection{First-order decoupling of the mode-matching equations}
\label{firstorderdecoupling}

The Schr\"{o}dinger equation $H\psi=E\psi$ produces 8 coupled differential equations, and an attempt at direct solution produces unwieldy results. Our approach is to partially decouple these by suitable unitary transformations of $H$. We take the inversion symmetry breaking strength $\lambda$ and chemical potential $\mu$ as small parameters and seek a decoupling up to corrections of first or second order in $\lambda,\mu$.

For a first-order decoupling we rotate the $\nu_x$ and $\tau_x$ spinors by the unitaries
\begin{equation}
U_\theta=\exp\left(\tfrac{1}{2}i\theta\nu_y\tau_z\sigma_z\right),\;\;U_\phi=\exp\left(\tfrac{1}{2}i\phi\nu_0\tau_y\sigma_z\right).\label{UthetaUphi}
\end{equation}
The rotation angles $\theta,\phi$ are $x$ and $k_z$-dependent,
\begin{subequations}
\label{firsttworotations}
\begin{align}
&\cos\theta=-(t_z/\Delta_{\rm eff})\sin k_z,\;\;\sin\theta=\Delta/\Delta_{\rm eff},\\
&\cos\phi=\Delta_{\rm eff}/M,\;\;\sin\phi=\tilde{m}/M,\\
&\Delta_{\rm eff}(x)=\sqrt{\Delta^2(x)+t_z^2\sin^2 k_z},\\
&M(x)=\sqrt{\tilde{m}^2+\Delta^2(x)+t_z^2\sin^2 k_z}.\label{UphiUthetadef}
\end{align}
\end{subequations}
Notice that $\cos\theta\rightarrow -{\rm sign}\,k_z$ for $\Delta\rightarrow 0$. We can avoid this discontinuity at $k_z=0$ by keeping a small nonzero $\Delta$ in the normal region.

The transformed Hamiltonian,
\begin{widetext}
\begin{align}
H_{\phi,\theta}={}&U_\phi^\dagger U_\theta^\dagger HU_\theta^{\vphantom{\dagger}} U_\phi^{\vphantom{\dagger}}\nonumber\\
={}&t\nu_z \tau_z (\sigma_x k_x + \sigma_y \sin k_y)- M \nu_z \tau_z \sigma_z+ \beta \nu_0 \tau_0 \sigma_z
+V_b(x)
  \nonumber\\
&- \mu \cos\theta \, \nu_z \tau_0 \sigma_0  - \mu\sin\theta  \cos\phi \, \nu_x \tau_z \sigma_z - \mu \sin\theta \sin\phi \, \nu_x \tau_x \sigma_0 \nonumber\\
&+\lambda\sin\theta \, \nu_x\tau_0\sigma_z+ \lambda \cos\theta  \cos\phi \, \nu_z \tau_z \sigma_0 + \lambda \cos\theta\sin\phi \, \nu_z\tau_x\sigma_z,\label{Htransformed}
\end{align}
\end{widetext}
is diagonal in the $\nu$ and $\tau$ degrees of freedom up to corrections of first order in $\lambda,\mu$, and up to a boundary potential $V_b(x)$ resulting from  the commutator of $k_x=-i\partial/\partial x$ and the $x$-dependent superconducting gap $\Delta(x)$ at the NS interface. In this section we discard the boundary potential, to simplify the calculations --- we will fully include it in the Appendix.

The term $\propto \mu\nu_x\tau_x\sigma_0$ in the Hamiltonian \eqref{Htransformed} can be made diagonal in $\nu$ and $\tau$ with the unitary transformation
\begin{subequations}
\label{thirdrotation}
\begin{align}
&H_{\psi,\phi,\theta}= U_\psi^\dagger P_3^\dagger H_{\phi,\theta}P_3 U_\psi,\\
&U_\psi=\exp(\tfrac{1}{2}i\psi\nu_0\tau_y\sigma_0),\\
&P_3=\frac{1}{2}\begin{pmatrix}
(\tau_0+\tau_z)\sigma_0&(\tau_x-i\tau_y)\sigma_0\\
(\tau_0-\tau_z)\sigma_0&(\tau_x+i\tau_y)\sigma_0
\end{pmatrix},\\
&\cos\psi=(1-\sin^2\theta\cos^2\phi)^{-1/2}\cos\theta,\\
&\sin\psi=-(1-\sin^2\theta\cos^2\phi)^{-1/2}\sin\phi\sin\theta.
\end{align}
\end{subequations}
The four blocks in the shift matrix $P_3$  [with $(P_3)^3=1$] refer to the $\nu$ degree of freedom. The transformed Hamiltonian is
\begin{subequations}
\label{Htransformed2}
\begin{align}
H_{\psi,\phi,\theta}={}&H_{\rm diag}+ \delta H_{\rm diag}+ \delta H_{\rm offdiag},\\
H_{\rm diag}={}&t\nu_z \tau_0(\sigma_x  k_x + \sigma_y \sin k_y)\nonumber\\
&- M \nu_z \tau_0 \sigma_z+ \beta \nu_0 \tau_0 \sigma_z,
\label{Hdiag}\\
 \delta H_{\rm diag}={}&- \mu(1-\Delta^2/M^2)^{1/2}  \nu_0 \tau_z \sigma_0\nonumber\\
 & - \lambda (t_z/M)\nu_z \tau_0 \sigma_0\sin k_z,\\
 \delta H_{\rm offdiag}={}&\mu(\Delta/M)  \nu_y \tau_y \sigma_z  +\lambda(M^2-\Delta^2)^{-1/2}\nonumber\\
 &\quad\times\left[\tilde{m}\nu_x\tau_z\sigma_z-(\Delta/M)t_z\nu_x\tau_x\sigma_z\sin k_z\right].
\end{align}
\end{subequations}
The symbol $\delta$ keeps track of the order in $\lambda,\mu$ of the diagonal (``diag'') and off-diagonal (``offdiag'') blocks.

\subsection{Second-order decoupling via Schrieffer-Wolff transformation}

The Schrieffer-Wolff transformation
\begin{align}
&H_{\rm SW}= e^{i\delta S}H_{\psi,\phi,\theta}e^{-i\delta S},\\
&\;\;\delta S=\begin{pmatrix}
0&\delta s\\
\delta s^\dagger&0
\end{pmatrix}\equiv \tfrac{1}{2}(\nu_x+i\nu_y)\delta s+\tfrac{1}{2}(\nu_x-i\nu_y)\delta s^\dagger,\nonumber
\end{align}
with Hermitian off-diagonal matrix $\delta S$ given by
\begin{equation}
[\delta S,H_{\rm diag}]=i\delta H_{\rm offdiag},\label{deltaSequation}
\end{equation}
removes the off-diagonal blocks up to corrections of second order in $\delta$:
\begin{align}
H_{\rm SW}={}&H_{\rm diag}+ \delta H_{\rm diag}+{\cal O}( \delta^2).
\end{align}
The solution of Eq.\ \eqref{deltaSequation} is \cite{Sylvester}
\begin{align}
\delta s=\frac{1}{2\beta M}\left[\frac{\lambda}{(M^2-\Delta^2)^{1/2}}\left(\tilde{m}\tau_z-\frac{\Delta t_z\sin k_z}{M}\tau_x\right) \right.\nonumber\\
\left.-\frac{\mu\Delta}{M}i\tau_y\right]\bigl(i\beta\sigma_0+\sigma_y tk_x-\sigma_x t\sin k_y\bigr).
\end{align}

The Schrieffer-Wolff matrix $\delta S$ contributes terms of order $\delta^2$ to the energy spectrum, which is given by the eigenvalues of $H_{\rm diag}+ \delta H_{\rm diag}+\delta H_{\rm SW}$ with
\begin{equation}
\delta H_{\rm SW}=\tfrac{1}{2}i[\delta S,\delta H_{\rm offdiag}]+i[\delta S,\delta H_{\rm diag}]+{\cal O}(\delta^3).\label{deltaHSW} 
\end{equation}

\subsection{Dispersion relation of the surface modes}
\label{modematchingcal}

The mode-matching calculation at energy $E$ with the Hamiltonian $H_{\rm diag}+\delta H_{\rm diag}$ (not yet including the Schrieffer-Wolff correction) now involves four uncoupled differential equations, labeled by $\nu,\tau\in\{-1,+1\}$, for a two-component spinor $\psi(x)$:
\begin{equation}
\begin{split}
&t\nu\frac{d\psi}{dx}=\bigl[i(E+{\cal U})\sigma_x+t\nu\sigma_z\sin k_y+(M\nu-\beta)\sigma_y]\psi\\
&{\cal U}=\mu\tau(1-\Delta^2/M^2)^{1/2}+\lambda (t_z/M)\nu\sin k_z.
\end{split}\label{diffequation}
\end{equation}
We solve this for piecewise constant coefficients. For the normal (N) region at $0<x<W$ we choose
\begin{subequations}
\label{piecewise}
\begin{equation}
\Delta=\Delta_{\rm N},\;\;\mu=\mu_{\rm N},\label{piecewisea}
\end{equation}
and for the superconducting (S) region at $x<0$ and $x>W$ we choose
\begin{equation}
\Delta=\Delta_{\rm S},\;\;\mu=\mu_{\rm S},\label{piecewiseb}
\end{equation}
\end{subequations}
demanding continuity of $\psi(x)$ at $x=0,W$. We keep a finite pair potential $\Delta_{\rm N}$ in the normal region to avoid the discontinuity at $p_z=0$ noted in Sec.\ \ref{firstorderdecoupling}.

To obtain the dispersion relation at a single NS interface we may take $W\rightarrow\infty$ and match decaying wave functions at both sides of the interface at $x=0$. Such a bound surface state is possible if $M\nu-\beta$ has the opposite sign in N and S, which requires $\nu=+1$ (since $\beta$ and $M$ are both positive). We denote $M\equiv  M_{\rm N}$ in N and $M\equiv M_{\rm S}$ in S, and similarly denote
\begin{equation}
\pm\mu(1-\Delta^2/M^2)^{1/2}+\lambda (t_z/M)\sin k_z\equiv\begin{cases}
{\cal U}_{\rm N}^\pm&\text{in N},\\
{\cal U}_{\rm S}^\pm&\text{in S}.
\end{cases}
\end{equation}
The sign $\pm$ accounts for the quantum number $\tau$ in Eq.\ \eqref{diffequation}. 

For a surface state we need $M_{\rm N}-\beta<-|{\cal U}_{\rm N}^\pm|$, $M_{\rm S}-\beta>|{\cal U}_{\rm S}^\pm|$ in some interval of $E,k_y,k_z$ around zero. Solution of Eq.\ \eqref{diffequation} gives the wave function profile
\begin{align}
&\psi(x)=C_{\rm N}e^{-x\kappa_{\rm N}^\pm/t}\begin{pmatrix}
i\kappa_{\rm N}^\pm-it\sin k_y\\
E+{\cal U}_{\rm N}^\pm+M_{\rm N}-\beta
\end{pmatrix},\;\;{\rm for}\;\;x>0,\\
&\psi(x)=C_{\rm S}e^{x\kappa_{\rm S}^\pm/t}\begin{pmatrix}
-i\kappa_{\rm S}^\pm-it\sin k_y\\
E+{\cal U}_{\rm S}^\pm+M_{\rm S}-\beta
\end{pmatrix},\;\;{\rm for}\;\;x<0,
\end{align}
with inverse decay lengths
\begin{equation}
\kappa_{\rm N,S}^\pm=\sqrt{t^2\sin^2 k_y+(M_{\rm N,S}-\beta)^2-(E+{\cal U}_{\rm N,S}^{\pm})^2}
\end{equation}
on the normal and superconducting sides of the NS interface.

The amplitudes $C_{\rm N}$ and $C_{\rm S}$ are to be adjusted so that $\psi(x)$ is continuous at $x=0$. By requiring that the matrix of coefficients of the mode-matching equations has vanishing determinant, we arrive at the dispersion relation of the surface modes,
\begin{align}
E_\pm(k_y,k_z)={}&
t\sin k_y
+\frac{(M_{\rm N}-\beta){\cal U}^\pm_{\rm S}-(M_{\rm S}-\beta){\cal U}^\pm_{\rm N}}{M_{\rm S}-M_{\rm N}}\nonumber\\
&\qquad\qquad+{\cal O}(\delta^2),\label{Eresult}
\end{align}
discarding terms of second order in $\mu,\lambda$. The level crossing at $k_z=0$, for a given $k_y$, happens for $m_0=t(\cos k_y-1)$. The corresponding charge expectation value $Q=-e\partial E/\partial\mu$ is
\begin{align}
Q_\pm=&\mp e(M_{\rm S}-M_{\rm N})^{-1}\biggl[(M_{\rm N}-\beta)\sqrt{1-\Delta^2_{\rm S}/M_{\rm S}^2}\nonumber\\
&-(M_{\rm S}-\beta)\sqrt{1-\Delta^2_{\rm N}/M_{\rm N}^2}\biggr]+{\cal O}(\delta),\label{Qresult}
\end{align}
one order in $\mu,\lambda$ less accurate than the energy. 

\begin{figure}[tb]
\centerline{\includegraphics[width=0.9\linewidth]{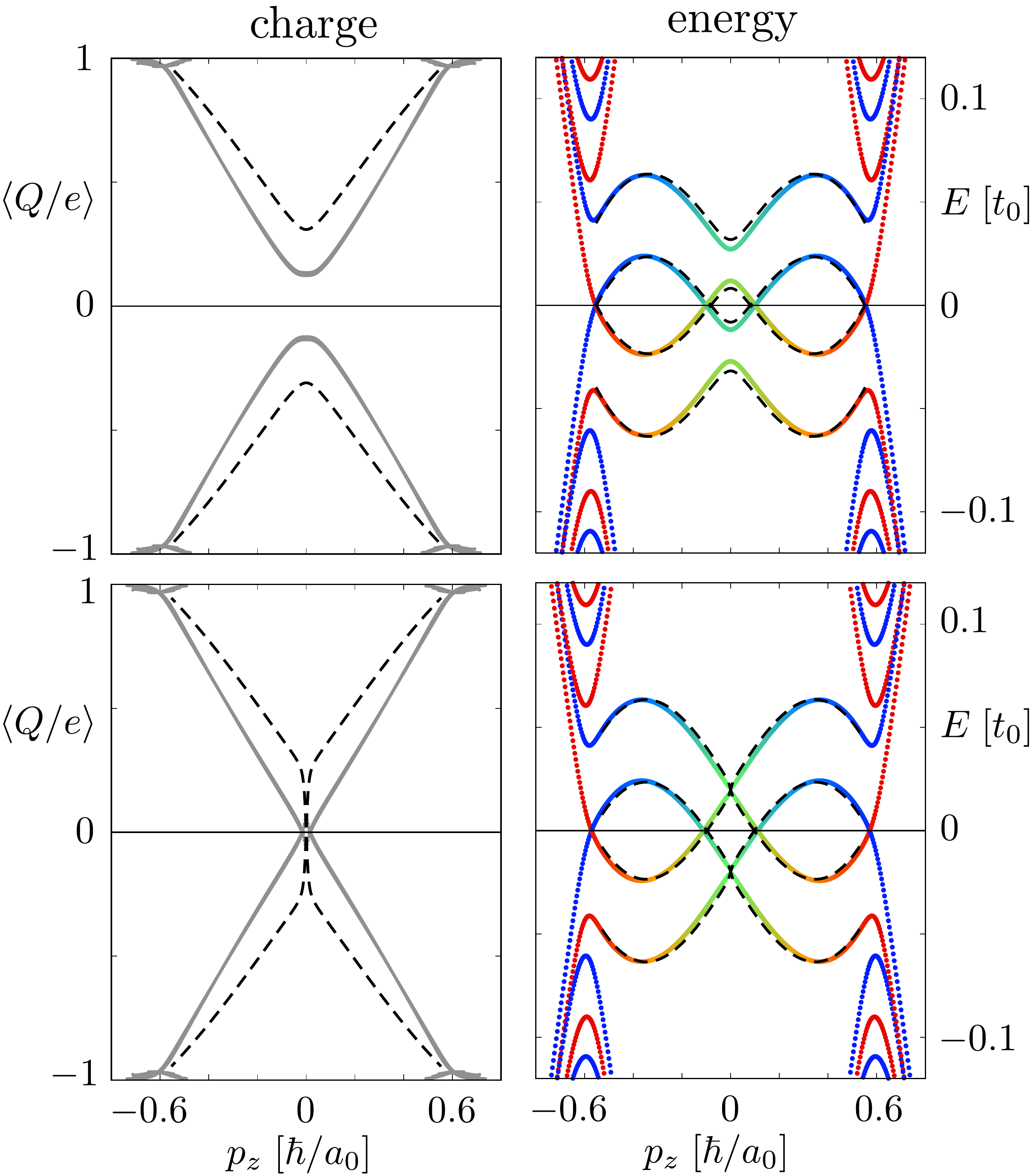}}   
\caption{Colored data points: Energy spectrum (color scale as in Fig. \ref{fig_numerics1}) and average charge obtained from a numerical diagonalization of the discretized Hamiltonian \eqref{Hx}. The top row is for $m_0=0.05$, the bottom row for $m_0=0$, other parameters: $t=2$, $t_z=1$, $\lambda=0$, $\beta=0.6$, $\mu_{\rm N}=\Delta_{\rm N}=10^{-2}$, $\mu_{\rm S}=0.2$, $\Delta_{\rm S}=0.8$, $W=120$, $k_y=0.01$. The black dashed curves result directly from the analytical mode-matching calculation, Eqs.\ \eqref{Eresult} and \eqref{Qresult}, \textit{without any adjustable parameters.}} 
\label{fig_modematching}
\end{figure}

In Fig.\ \ref{fig_modematching} we compare the numerical diagonalization of the Hamiltonian \eqref{Hx} with the analytical mode matching calculation. Unlike the comparison in Fig.\ \ref{fig_numerics_2}, here there is not a single fit parameter. The agreement is excellent for the energy, somewhat less for the average charge.  

\subsection{Effective surface Hamiltonian}
\label{sec_effH}

In Sec.\ \ref{sec_effectiveH} we constructed an effective surface Hamiltonian by relying only on particle-hole symmetry. As an alternative route, we present here a derivation starting from the model Hamiltonian \eqref{Htransformed2}.

The motion perpendicular to the NS interface at $x=0$ is governed by the reduced Hamiltonian
\begin{equation}
H_{\perp}=t\nu_z \tau_0\sigma_x  k_x- M \nu_z \tau_0 \sigma_z+ \beta \nu_0 \tau_0 \sigma_z,\label{Hperp}
\end{equation}
with neglect of the terms $\propto\mu,\lambda$ as well as the $k_y$ and $k_z$-dependent terms for motion parallel to the interface. The wave function profile $\psi(x)$ at $E=0$,
\begin{align}
&H_\perp\psi=0\Rightarrow\psi(x)=\\
&\quad\exp\left[t^{-1}\int_0^x dx' \bigl(M(x')\nu_0\tau_0\sigma_y-\beta\nu_z\tau_0\sigma_y\bigr)\right]\psi(0),\nonumber
\end{align}
decays for $x\rightarrow -\infty$ (inside the superconducting region) because of the term $\propto M(-\infty)>\beta$ and for $x\rightarrow +\infty$ (inside the Weyl semimetal region) because of the term $\propto\beta>M(\infty)$. This two-sided decay is ensured if $\psi(0)$ is an eigenstate with eigenvalue $+1$ of both $\nu_0\tau_0\sigma_y$ and $\nu_z\tau_0\sigma_y$. The resulting eigenspace has rank two.

The $2\times 2$ effective surface Hamiltonian $H_{\rm eff}$ for motion parallel to the surface is obtained by projecting $H$ onto this two-dimensional eigenspace, resulting in
\begin{align}
H_{\rm eff}
={}&  \tau_0 t\sin k_y-\lambda(t_z/M)\tau_0 \sin k_z\nonumber\\
& -\mu(1-\Delta^2/M^2)^{1/2}\tau_z  .\label{Heff}
\end{align}
The corresponding charge operator is momentum dependent,
\begin{equation}
Q_{\rm eff}=-e\,\partial H_{\rm eff}/\partial\mu=e(1-\Delta^2/M^2)^{1/2}\tau_z.
\end{equation}
In this effective surface description the energy scales $\Delta$ and $\mu$ should be regarded as weighted averages of the $x$-dependent parameters from Eq.\ \eqref{piecewise}.

The two surface modes have opposite charge $Q_\pm=\pm e\,(1-\Delta^2/M^2)^{1/2}$ and dispersion relation
\begin{align}
&E_{\pm}(k_z)=t\sin k_y-(\Delta^2+\tilde{m}^2(k_y,k_z)+t_z^2\sin^2 k_z)^{-1/2}\nonumber\\
&\qquad\times\left[\lambda t_z\sin k_z\pm \mu\sqrt{\tilde{m}^2(k_y,k_z)+  t_z^2\sin^2 k_z}\right],\label{Epmresult}
\end{align}
representing the spiraling surface Fermi arc illustrated in Fig.\ \ref{fig_layout}. The $\pm$ index corresponds to the $\nu$ index of Sec.\ \ref{sec_effectiveH}, the $\kappa$ index is taken care of by the sign of $\sin k_y$. The gap $\delta E=E_+(0)-E_-(0)$ at $k_z=0$ equals
\begin{equation}
\delta E=\frac{2\mu m_{\rm eff}}{\sqrt{m_{\rm eff}^2+\Delta^2}},\;\;m_{\rm eff}=|m_0+t(1-\cos k_y)|.\label{meffdef}
\end{equation}
We interpret $m_{\rm eff}$ as the effective coupling strength of the surface state to the superconductor, and as the parameter that in the microscopic model of Sec.\ \ref{sec_micro} is varied by varying $U_{\rm barrier}$. The level crossing then happens when $m_{\rm eff}=0$. At the level crossing the excitations are charge neutral.

We may include the Schrieffer-Wolff correction, by projecting $\delta H_{\rm SW}$ from Eq.\ \eqref{deltaHSW} onto the surface eigenspace. The result is a correction of order $\delta^2$ to the effective surface Hamiltonian,
\begin{align}
\delta H_{\rm eff}=
&-\frac{t\sin k_y}{2\beta M^3}\biggl(2\mu\lambda \frac{\Delta^2\tau_z t_z\sin k_z +\Delta M\tilde{m}\tau_x}{\sqrt{M^2-\Delta^2}}\nonumber\\
&+(\lambda^2\tilde{m}^2+\lambda^2\Delta^2+\mu^2\Delta^2)\tau_0\biggr).
\end{align}
The dominant effect of this correction is to shift the level crossing away from $k_z=0$ to $k_z=-(\lambda/\beta)(t/t_z)\sin k_y$.

\section{Conclusion}
\label{sec_conclude}

In summary, we have investigated the superconducting proximity effect on the dispersion relation of surface modes in a Weyl-Majorana solenoid --- a Weyl semimetal nanowire with an axial magnetization covered by a superconductor. The surface Fermi arc connecting bulk Weyl cones is broken up into nearly charge-neutral Majorana modes. We have identified a ``connectivity index'' that determines between which pair of modes a gap is opened by the superconductor. 

We have discovered that the sign of the induced gap can be inverted by variation of the tunnel coupling strength between the semimetal and the superconductor. A domain wall separating segments of the nanowire with opposite sign of the gap traps a charge-neutral quasiparticle. This bound Majorana fermion is not at zero energy, so it should not be confused with the Majorana zero-modes in semiconductor nanowires \cite{Bee13,Ell15,Sat16}. The gap inversion is studied for a 3D model Hamiltonian, both numerically in a tight-binding formulation, and analytically via mode matching at the normal-superconductor interface. Further insight is obtained by an effective 2D surface Hamiltonian.

In closing we remark on a global aspect of the gap inversion in terms of the flow of Berry curvature (topological charge) in the Brillouin zone \cite{Mur03}. The minimal number of two Weyl cones in a Weyl semimetal with broken time-reversal symmetry is doubled if we include the electron-hole degree of freedom. The sign of the Berry curvature at a given point in the Brillouin zone is not changed by the doubling \cite{Men12}, so the Fermi arc connecting Weyl cones of opposite Berry curvature must still run across the Brillouin zone --- but now it has a choice: it may connect cones of the same or opposite electrical charge. If we inspect Fig.\ \ref{fig_numerics_2} we see that the Fermi arcs always connect Weyl cones of the same electrical charge (coded blue or red), except at the gap inversion point. At the critical tunnel barrier height $U_{\rm barrier}=U_c$ the Majorana surface modes connect bulk states of opposite electrical charge (from blue to red).

In Fig.\ \ref{fig_numerics_2} the anomalous connection by Fermi arcs of Weyl cones of opposite electrical charge and opposite topological charge happens only at an isolated point in parameter space, because the superconductivity is induced only at the surface of the Weyl semimetal. By inducing superconductivity throughout the bulk (for example, using the heterostructure approach of Ref.\ \onlinecite{Men12}) one should be able to stabilize the anomalous connection in an entire region of parameter space. We expect an anomalous Josephson effect to develop in the Weyl-Majorana solenoid as a result of this topologically nontrivial connection.

\acknowledgments

We have benefited from discussions with A. R. Akhmerov, C. L. Kane, T. Neupert, and T. E. O'Brien.
This research was supported by the Foundation for Fundamental Research on Matter (FOM), the Netherlands Organization for Scientific Research (NWO/OCW), and an ERC Synergy Grant.

\appendix

\section{Effect of the boundary potential on the mode-matching calculation}

The unitary transformations in Sec.\ \ref{analytics} introduce a boundary potential in the Hamiltonian \eqref{Htransformed2}, given by
\begin{widetext}
\begin{align}
V_{b}(x)={}& -it U_\psi^\dagger(x) P_3^\dagger U_\phi^\dagger(x) U_\theta^\dagger (x)\nu_z\tau_z\sigma_x\left[ \frac{\partial}{\partial x} ,U_\theta^{\vphantom{\dagger}}(x) U_\phi^{\vphantom{\dagger}}(x)P_3 U_\psi(x)\right]\nonumber\\
={}&\tfrac{1}{2}t(\theta'\sin\phi+\psi')\nu_z\tau_y\sigma_x
-\tfrac{1}{2}t (\phi' \sin \psi+\theta' \cos \psi \cos \phi)\nu_x\tau_x\sigma_y
-\tfrac{1}{2}t (\phi' \cos \psi-\theta' \sin \psi \cos \phi)\nu_x \tau_z\sigma_y\nonumber\\
={}&-\frac{\tfrac{1}{2}tm_z}{\Delta^2(x)+m_z^2}\,\frac{d\Delta(x)}{dx}\nu_x\tau_x\sigma_y,
\end{align}
\end{widetext}
where we abbreviated
\begin{equation}
m_{z}=(\tilde{m}^2+t_z^2\sin^2 k_z)^{1/2}.
\end{equation}
For simplicity we omitted $V_b(x)$ from the mode-matching calculations and the derivation of the effective surface Hamiltonian in Sec.\ \ref{analytics}. In the following we include it in the calculation, resulting in an improved agreement of the analytics with the numerics but without simple closed-form expressions as Eqs.\ \eqref{Eresult} and \eqref{Qresult}.

The step-function variation of the pair potential $\Delta(x)$ at the NS interfaces $x=0,W$ produces a delta-function boundary potential. Let us focus on the interface at $x=0$, with $\Delta=\Delta_{\rm N}$ for $x>0$ and $\Delta=\Delta_{\rm S}$ for $x<0$. Because of the boundary potential, the wave function does not vary continuously across the NS interface. Instead, the wave functions at the two sides of the interface $x=0$ are related by the transfer matrix, 
\begin{equation}
\begin{split}
&\psi(0^+)=e^{i{\cal M}_{\rm NS}}\psi(0^-),\\
&{\cal M}_{\rm NS}=-\frac{1}{t}\int_{0^-}^{0^+} dx\, \nu_z\tau_0\sigma_x V_{b}(x)=-\tfrac{1}{2}\alpha\nu_y\tau_x\sigma_z,
\end{split}\label{MNS}
\end{equation}
where the angle $\alpha$ is given by the integral
\begin{equation}
\alpha=\int_{\Delta_{\rm S}}^{\Delta_{\rm N}}d\Delta\,\frac{m_z}{\Delta^2+m_z^2}={\rm arctan}\,\frac{\Delta_{\rm N}}{m_z}-{\rm arctan}\,\frac{\Delta_{\rm S}}{m_z}.
\end{equation}
Note that at the level crossing point we have $m_z=  0$ hence $\alpha=  0$, so the level crossing itself is not affected by the boundary potential.

As explained in Sec.\ \ref{sec_effH}, to obtain the effective surface Hamiltonian we impose a two-sided decay of the wave function, by demanding that $\psi$ is an eigenstate with eigenvalue $+1$ of $\nu_0\tau_0\sigma_y$ in S and of $\nu_z\tau_0\sigma_y$ in N. The former condition can be rewritten as a boundary condition in N,
\begin{equation}
\psi(0^+)=U_b\psi(0^+),\;\;U_b=e^{i{\cal M}_{\rm NS}}\nu_0\tau_0\sigma_y e^{-i{\cal M}_{\rm NS}}.
\end{equation}
Note that $U_b$ and $\nu_z\tau_0\sigma_y$ commute, so they can be diagonalized simultaneously. 
The rank two eigenspace of eigenvalue $+1$ is spanned by the vectors
\begin{align}
&v_1=\left(0,0,\sin \alpha,i \sin \alpha,1-\cos \alpha,-i +i\cos \alpha,0,0\right),\nonumber \\
&v_2=\left(\sin \alpha,i \sin \alpha,0,0,0,0,1-\cos \alpha,-i +i\cos \alpha\right).\nonumber
\end{align}
The Hamiltonian projected onto this eigenspace is
\begin{equation}
\begin{split}
&H_{\rm eff}=\tau_0 t\sin k_y-(\gamma /\bar{M})(\lambda\tau_0 t_z\sin k_z -\mu \tau_zm_z),\\
&\gamma=\cos\alpha+(\bar{\Delta}/m_z)\sin\alpha,
\end{split}
\end{equation}
where the $x$-dependent gap $\Delta(x)$ in the full Hamiltonian has been replaced by a spatial average $\bar{\Delta}$, and $\bar{M}=(m_z^2+\bar{\Delta}^2)^{1/2}$.

Comparison with Eq.\ \eqref{Heff} shows that the effect of the boundary potential is to renormalize the parameters $\lambda$ and $\mu$ by a factor $\gamma$. For $\Delta_{\rm S}\gg m_z$ we have
\begin{equation}
\gamma=(\Delta_{\rm N}^2+m_z^2)^{-1/2}(\Delta_{\rm N}-\bar{\Delta}).
\end{equation}

\begin{figure}[tb]
\centerline{\includegraphics[width=0.9\linewidth]{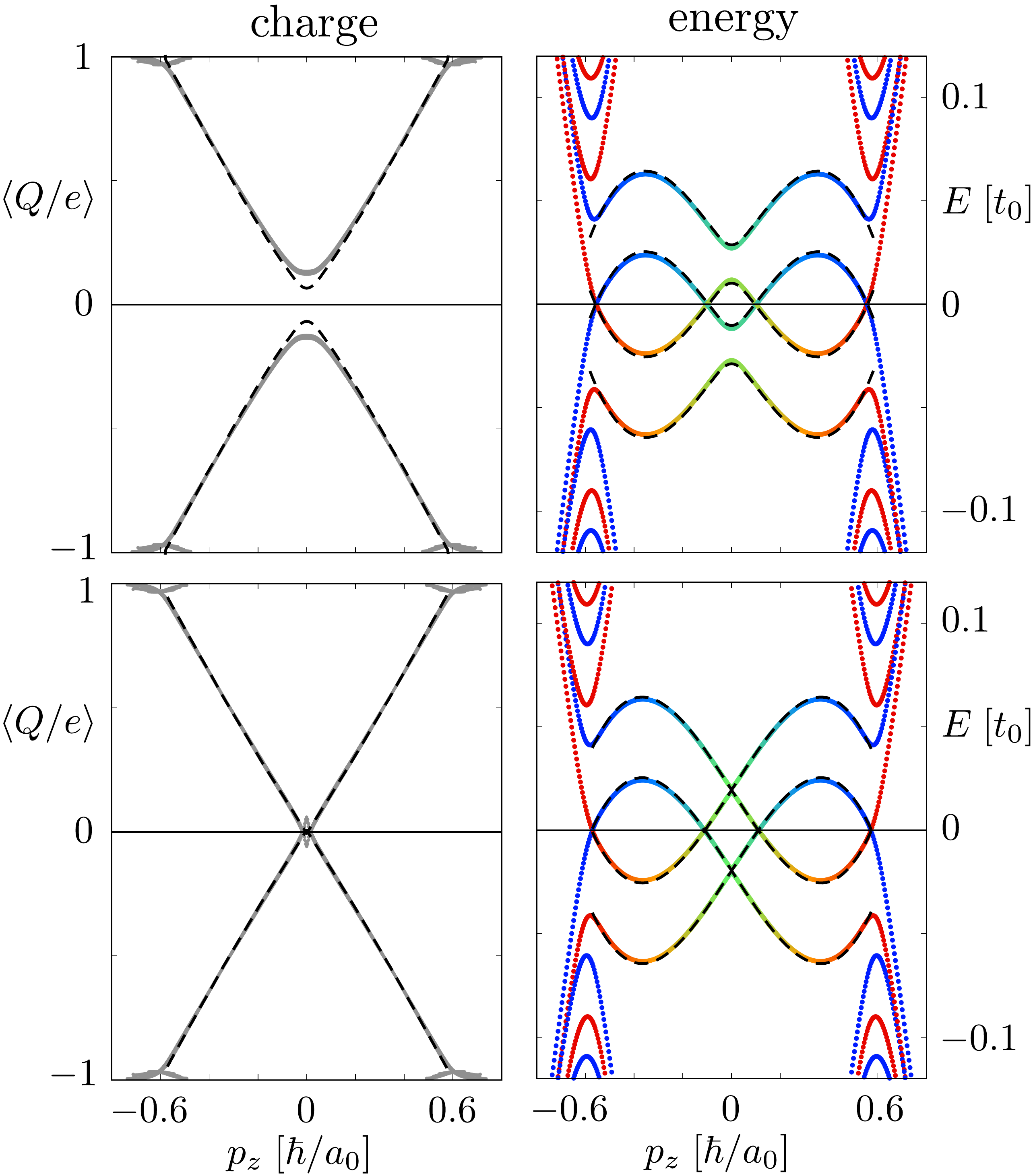}}   
\caption{Colored data points: Energy spectrum (color scale as in
  Fig. \ref{fig_numerics1}) and average charge obtained from a
  numerical diagonalization of the discretized Hamiltonian
  \eqref{Hx}. The parameters are the same as in Fig.\
  \ref{fig_modematching}. The black dashed curves result from the
  mode-matching calculations including the
  boundary potential and the full Hamiltonian (with the off-diagonal terms).}
\label{fig_modematchingcorr}
\end{figure}

The full mode-matching calculation of Sec.\  \ref{modematchingcal} is also modified by the new boundary condition. Since Eq.\ \eqref{MNS} mixes the $\nu$ and $\tau$ indices, we can no longer use the block-diagonalization of the Hamiltonian to simplify the mode matching, and we could not find a closed-form solution analogous to Eqs.\ \eqref{Eresult} and \eqref{Qresult}. Including both the diagonal and off-diagonal terms in the Hamiltonian \eqref{Htransformed2} we find the energy and charge expectation value shown in Fig.\ \ref{fig_modematchingcorr} (dashed curves). The solid curves are the numerical solution of the tight-binding model. Comparison with Fig.\ \ref{fig_modematching}, where we did not include the boundary potential and discarded off-diagonal $\nu,\tau$ terms in the Hamiltonian, shows little difference in the energy but an improved agreement in the charge.


\begin{thebibliography}{99}
\bibitem{parameters_num1} The microscopic model parameters in the slab geometry of Fig.\ \ref{fig_numerics1} are (energies in units of $t_0$, lengths in units of $a_0$): $t=2$, $t_z=1$, ${m}_0=-0.3$, $\lambda=0$, $\beta=0.6$, $\tilde{t}=7$, $\tilde{t}_z=3.5$, $\mu_{\rm W}=10^{-4}$, $\mu_{\rm S}=3.5$, $U_{\rm barrier}=0.1$, $\Delta_0=0.2$, $d_{\rm barrier}=2$, $W=120$, $k_y=\pi/120$.
\bibitem{parameters_fit} The fit parameters used in Fig. \ref{fig_numerics_2} are $D_0(p_z)= 0.016 + 0.035\,p_z^2 + 0.24\, p_z^4$, $c= 0.085$, $c'' = 0.056$.
\bibitem{note1} The slab geometry has a $\pm k_y$ degeneracy in the spectrum, corresponding to surface states at the opposite NS interfaces $x=0,W$. We therefore only need to show a single sign of $k_y$ to obtain the full spectrum, as in Fig.\ \ref{fig_numerics1}.
\bibitem{note_EF} In figures \ref{fig_numerics1}, \ref{fig_numerics_2}, and \ref{fig_trapping} we also added a small offset of $10^{-4}\,t_0$ to $\mu_{\rm W}$ to break the electron-hole degeneracy.
\bibitem{parameters_num4} The microscopic model parameters in the wire geometry of Fig.\ \ref{fig_numerics_4} are (energies in units of $t_0$, lengths in units of $a_0$): $t=2$, $t_z=1$, ${m}_0=-0.3$, $\lambda=0$, $\beta=0.6$, $\tilde{t}=7$, $\tilde{t}_z=3.5$, $\mu_{\rm W}=0.05$, $\mu_{\rm S}=3.5$, $\Delta_0=0.4$, $d_{\rm barrier}=1$, $W=79$.
\bibitem{Sylvester} To solve Eq.\ \eqref{deltaSequation} for $\delta s$ we substitute the block-decomposition $H_{\rm diag}=\begin{pmatrix}
h_+&0\\
0&h_-
\end{pmatrix}$, $\delta H_{\rm offdiag}=\begin{pmatrix}
0&\delta h\\ 
\delta h^\dagger&0
\end{pmatrix}$ of the $8\times 8$ matrices $H_{\rm diag}$ and $\delta H_{\rm offdiag}$ in the $\nu$ degree of freedom. We thus arrive at the equation $\delta s h_- -h_+\delta s=i\delta h$ involving $4\times 4$ matrices. This Sylvester equation has a unique solution (unless $h_+$ and $h_-$ have a common eigenvalue, which they do not).
\bibitem{noteNSvsSNS} Fig.\ \ref{fig_trapping} is for an NS junction with a single interface at positive and negative $k_y=\pm \pi/120$. Figs.\ \ref{fig_numerics1}, \ref{fig_numerics_2}, and \ref{fig_numerics_3} are for an SNS junction with two interfaces at a single $k_y=\pi/120$.
\end{thebibliography}
\end{document}